\documentclass[onecolumn]{aastex62}

\pdfoutput=1

\usepackage[fleqn]{amsmath}
\usepackage{graphicx}

\usepackage{subfiles}
\usepackage{multirow}

\usepackage{savesym}
\savesymbol{tablenum}
\usepackage[separate-uncertainty=true]{siunitx}
\restoresymbol{SIX}{tablenum}

\renewcommand{\Re}{\operatorname{Re}}
\renewcommand{\Im}{\operatorname{Im}}

\submitjournal{ApJ}
\received{01/29/2019}
\accepted{03/01/2019}

\shorttitle{Tides between the TRAPPIST-1 planets}
\shortauthors{Hay and Matsuyama}

\begin{document}

\title{Tides between the TRAPPIST-1 planets}

\correspondingauthor{Hamish Hay}
\email{hhay@lpl.arizona.edu}

\author[0000-0002-0786-7307]{Hamish C. F. C. Hay}
\affil{Lunar and Planetary Laboratory, University of Arizona, Tucson, AZ 85719, USA}

\author{Isamu Matsuyama}
\affiliation{Lunar and Planetary Laboratory, University of Arizona, Tucson, AZ 85719, USA}


\begin{abstract}

The TRAPPIST-1 system is sufficiently closely packed that tides raised by one planet on another are significant. We investigate whether this source of tidal heating is comparable to eccentricity tides raised by the star. Assuming a homogeneous body with a Maxwell rheology, we find that energy dissipation from stellar tides always dominates over that from planet-planet tides across a range of viscosities. TRAPPIST-1 g may experience the greatest proportion of planet-planet tidal heating, where it can account for between \SI{2}{\percent} and \SI{20}{\percent} of the total amount of tidal heating, for high (\SI{e21}{\pascal\second}) and low viscosity (\SI{e14}{\pascal\second}) regimes, respectively. If planet-planet tidal heating is to exceed that from stellar eccentricity tides, orbital eccentricities must be no more than $e=$ \numrange{e-3}{e-4} for most of the TRAPPIST-1 planets.

\end{abstract}


\section{Introduction} \label{sec:intro}

Tidal heating occurs due to internal friction as a body deforms in response to a time-varying external gravitational potential. It is known to be a dominant process for a number of solar system bodies, such as the Jovian moon Io \citep[e.g.,][]{peale1979melting}, the small Saturnian satellite Enceladus \citep[e.g.,][]{squyres1983evolution,ross1989viscoelastic, roberts2008tidal}, Triton, a retrograde satellite of Neptune \citep{nimmo2015powering}, and Earth. Indeed, tidal heating in the Galilean satellites is likely the reason they remain in a stable orbital configuration today.

We usually consider tides raised by the central body on the orbiting body and vice versa. Orbital eccentricity of the secondary object causes it to pass through a time-varying tidal potential, which induces heating due to periodic deformation. Yet, other secondary objects in a system are also sources of time-varying tidal forces. Such tides are typically negligible because the mass of the central tide raising body is usually far greater than other bodies in the system, and also because the distances between these bodies are vast and the strength of tidal forces decreases with the distance between them cubed. The seven planet extrasolar system, TRAPPIST-1 \citep{gillon2016,gillon2017}, is the first system to be discovered where this is not the case. The separation distance at conjunction is small enough that tides raised by neighbouring planets can become significant, and heating must occur as a result. While tidal heating due to orbital eccentricity has been investigated for the TRAPPIST-1 system \citep[e.g.,][]{barr2018interior}, it has never been addressed in detail for planet-planet tides \citep{wright2018}.

There are two main aims of this paper. First, we provide the community with the first theory of planet-planet tides and associated heating (Section \ref{sec:theory}). Second, we apply our theory to the TRAPPIST-1 planets by assuming homogeneous interior structures and Maxwellian rheology (Section \ref{sec:rheology}) to gain a first order understanding of the effect of planet-planet tides in the system and, in particular, how it compares to tidal heating from orbital eccentricity (Section \ref{sec:results}). 

\section{Tidal theory} \label{sec:theory}

The distance to a perturbing object, $r$, and the angle $\gamma$ between a position vector $\vec{P}$ on a planet's surface and the direction from the planet's centre to the perturber, are two fundamental properties that determine tidal forces acting on a planet. In this section we derive these for tides raised on one planet by another. First, we revisit these for tides raised by a central object.

The tide raising potential due to any object of mass $m$ at distance $r$ from a planet of radius $R$, limited to spherical harmonic degree-2,  is;

\begin{equation}\label{eq:tidal_potential_all}
\Phi^T = \frac{Gm}{r}\left(\frac{R}{r}\right)^2\left(\frac{3 \cos^2 \gamma - 1}{2}\right)
\end{equation}

\noindent where $\gamma$ is the angle between a point on the planet's surface and the line of centres between the two bodies \citep{murray1999solar}. This expression is sufficiently general that it can be used for tides due to a central object, or a neighbouring planet. The only difference between these two cases is how $r$ and $\gamma$ are derived. In the following we use the superscripts $e$ and $p$ to denote properties of eccentricity- and planet-forced tides, respectively.

\subsection{Tides raised by the central object}

In the reference frame of a planet orbiting a more massive central body and assuming eccentricity $e \ll 1$, the separation distance between two bodies is

\begin{equation}\label{eq:r_vec_1}
r \approx a \left(1 + e \cos M\right)^{-1}
\end{equation} 

\noindent to first order in eccentricity \citep{murray1999solar}. Here, $M=nt$ is the mean anomaly of the planet, $n$ is the mean motion, and $t$ is time. We assume that the planet is synchronously rotating, such that its rotation rate is equal to its mean motion. 

It can be shown that $\cos\gamma$ is related to the colatitude $\theta$ and longitude $\phi$ of a perturbed planet with zero obliquity via;

\begin{equation}\label{eq:cos_gamma_1}
\cos \gamma = \sin \theta \cos (\phi - \phi_t).
\end{equation}

\noindent where $\phi_t$ is a libration angle of the perturbing body's position in longitude that arises from small changes in the planet's orbital speed due to it's eccentric orbit. To first order in eccentricity, $\phi_t \sim 2 e \sin M$ \citep{murray1999solar}. Substituting this along with Eqs. \ref{eq:cos_gamma_1} and \ref{eq:r_vec_1} into Equation \ref{eq:tidal_potential_all} and ignoring the time-independent terms, we arrive at the dynamic tidal potential due to orbital eccentricity \citep[e.g.,][]{kaula1964tidal, wahr2006tides}:

\begin{equation}\label{eq:pot_tides_ecc}
\Phi^{e}(\theta, \phi, t) =
\frac{3}{2}\frac{Gm_{\star}}{a} \left(\frac{R}{a}\right)^2 e
\left[2 \sin M \sin^2 \theta \sin(2\phi) + \cos M \left( 3 \sin^2 \theta \cos^2 \phi -1\right) \right]
\end{equation} 

\noindent where the superscript $e$ denotes an eccentricity-forced potential. This time-varying tidal potential vanishes when $e = 0$. Dynamic tides due to neighbouring planets, which we consider in the next section, are unique in that they exist even for circular orbits.

\subsection{Tides raised by neighboring planets}

In this section we consider the situation shown in Figure \ref{fig:diagram}, where tides are raised on an inner body, planet $i$, by an outer body, planet $j$. Both planets orbit around the host star with coplanar, circular orbits, and have no obliquity. In the barycentric reference frame of the whole system, the position vectors of planets $i$ and $j$ are;

\begin{align}
\vec{r}_i &= a_i \left[\cos (n_i t) \hat{x} + \sin (n_i t) \hat{y} \right]\\
\vec{r}_j &= a_j \left[\cos (n_j t) \hat{x} + \sin (n_j t) \hat{y}\right]
\end{align}

where $\hat{x}$ and $\hat{y}$ are the inertial unit vectors in the $x$ and $y$ directions (Figure 1). The vector between the two planets is then the difference in these:

\begin{equation}\label{eq:r_vec_ij}
\vec{r}_{ij} = \vec{r}_j - \vec{r}_i = \left[ a_j \cos (n_j t) - a_i \cos (n_i  t ) \right] \hat{x} + \left[ a_j \sin (n_j t) - a_i \sin (n_i t) \right] \hat{y}.
\end{equation}

Taking the magnitude of this expression gives us the planet-planet separation distance as a function of time, a fundamental quantity in determining the magnitude of tidal forces;

\begin{equation}
r_{ij} = \left[a_i^2 + a_j^2 - 2 a_i a_j \cos(n_{ij} t)\right]^{1/2}
\end{equation}

\noindent where $n_{ij} = n_i - n_j$ is the conjunction/closest approach frequency between planets $i$ and $j$. As discussed above, we assume the planet is in synchronous rotation, and define planet $i$'s corotating unit vectors $\hat{x}_r$ and $\hat{y}_r$ to point towards the star and along the trailing hemisphere, respectively (Figure \ref{fig:diagram}). This corotating reference frame is related to the inertial frame via;

\begin{equation}\label{eq:rot_inert}
\begin{pmatrix}
\hat{x}  \\
\hat{y}
\end{pmatrix} =
\begin{pmatrix}
-\cos f_i & \sin f_i  \\
-\sin f_i & -\cos f_i
\end{pmatrix} 
\begin{pmatrix}
\hat{x}_r  \\
\hat{y}_r
\end{pmatrix}.
\end{equation}
 
\noindent where $f_i$ is the true anomaly of planet $i$. As we assume zero inclination and zero obliquity, $\hat{z} = \hat{z}_r$. In a similar fashion to eccentricity-forced tides (Eq. \ref{eq:cos_gamma_1}), $\cos \gamma$ can be expressed as;

\begin{equation}
\cos \gamma = \sin \theta \cos(\phi - \phi_{ij})
\end{equation} 

\noindent where now $\phi_{ij}$ is the angle between $\hat{x}_r$ and the planet-planet vector $\vec{r}_{ij}$ \eqref{eq:r_vec_ij}. For eccentricity-forced  tides, $\phi_t$ remains very small over the entire orbital period, whereas $\phi_{ij}$ varies from $0$ to $2 \pi$ for planet-forced tides, preventing us from making any small angle approximation. Rewriting $\hat{x}_r$ in the inertial reference frame using Eq. \ref{eq:rot_inert}, then taking its dot product with Eq. \ref{eq:r_vec_ij} and rearranging gives the following relationships for $\phi_{ij}$:

\begin{figure}[t]
	\centering
	\includegraphics[width=0.4\linewidth]{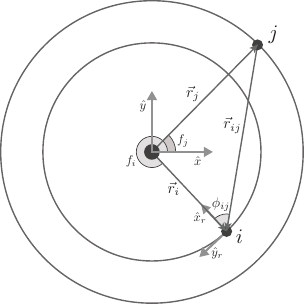}
	\caption{\label{fig:diagram} Schematic of the relevant properties for planet-planet tides. Both planets $i$ and $j$ orbit in the same plane around the central object. The inertial units vectors $\hat{x}$ and $\hat{y}$ are defined as shown. The corotating reference frame is shown for the inner planet, where $\hat{y}_r$ always points along the trailing hemisphere and $\hat{x}_r$ is directed towards the star. The stellar-planet vector is $\vec{r}_i$ and $\vec{r}_j$ for planets $i$ and $j$, respectively. The vector between the two planets is $\vec{r}_{ij} = \vec{r}_j - \vec{r}_i$. The angle between the stellar-planet vector and $\hat{x}$ is the true anomaly, $f$, of the planet.}
\end{figure}

\begin{align}
\cos \phi_{ij} &= \frac{a_i - a_j \cos (n_{ij}t)}{r_{ij}} \label{eq:cos_phi}\\
\sin \phi_{ij} &= \frac{a_j \sin (n_{ij}t)}{r_{ij}} \label{eq:sin_phi}
\end{align}

\noindent We can then rewrite $\cos \gamma$ in terms of Equations \ref{eq:cos_phi} and \ref{eq:sin_phi}:

\begin{equation}
\cos\gamma = \sin \theta 
\left(\cos \phi \cos \phi_{ij} + \sin \phi \sin \phi_{ij} \right)
\end{equation}

\noindent Substituting this expression into Equation \ref{eq:tidal_potential_all} gives the tidal potential on planet $i$ due to an outer planet $j$:

\begin{align}
\Phi^{p}_{ij} (\theta, \phi, t) &= \frac{1}{2}\frac{Gm_j}{r_{ij}} \left(\frac{R_i}{r_{ij}}\right)^2
\left[3 \sin^2 \theta 
\left(\cos \phi \cos \phi_{ij} + \sin \phi \sin \phi_{ij} \right)^2 - 1 \right]\, \text{for}\, i \neq j\label{eq:pot_pp}
\end{align}

\begin{figure}[t]
	\centering
	\includegraphics[width=0.5\linewidth]{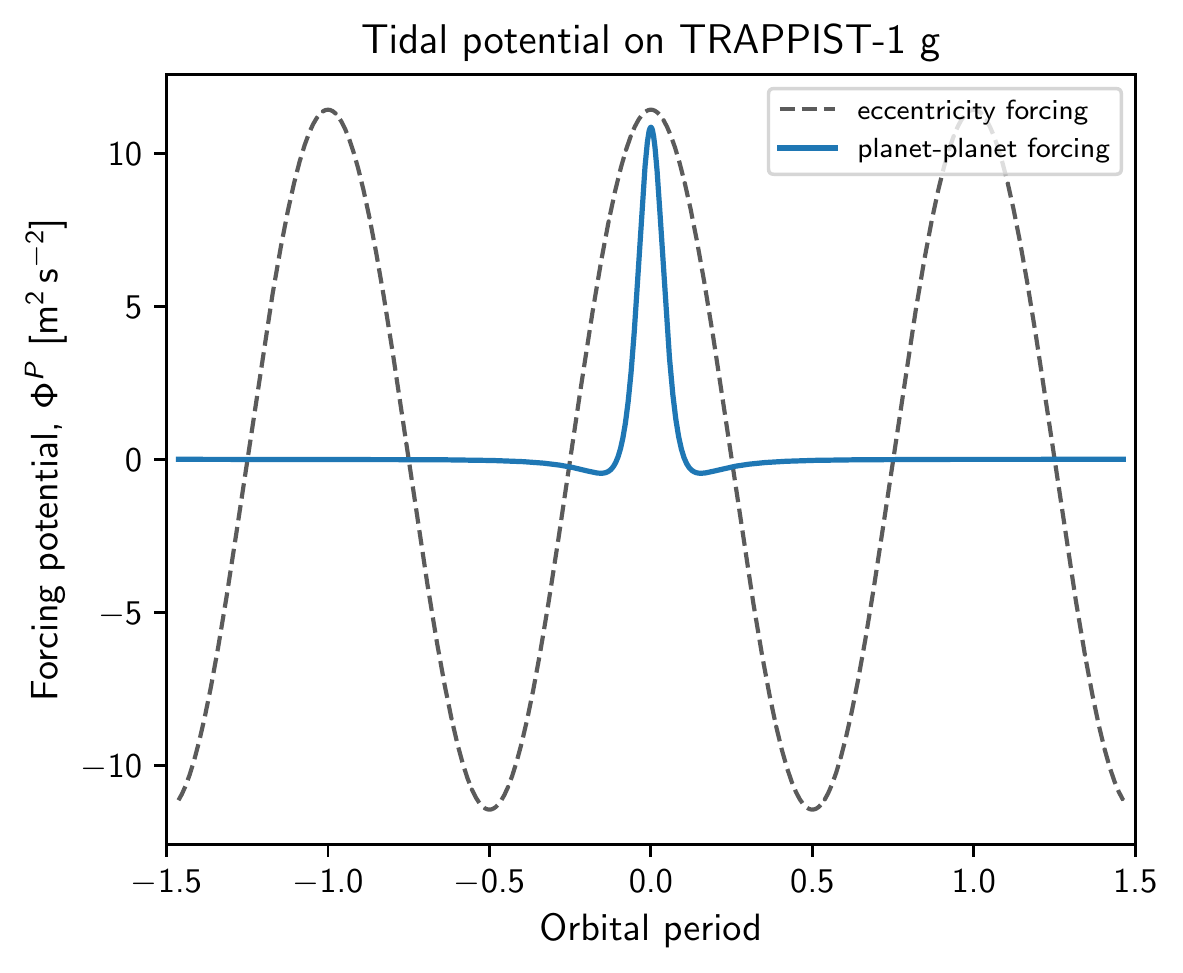}
	\caption{\label{fig:forcing_g} Amplitude of the tidal forcing potential at 0 degrees latitude and longitude on TRAPPIST-1 g, due to eccentricity (dashed line, Eq. \ref{eq:pot_tides_ecc}) and planet (blue line, Eq. \ref{eq:pot_pp}) tides, as a function of the planet's orbital period. The planet-planet forcing potential shown here is only due to TRAPPIST-1 f, using the values in Table \ref{tb:params2}.}
\end{figure}

\noindent where the superscript $p$ represents tidal forcing due to another planet. The above expression is sufficiently general that tides raised on the outer planet $j$ by the inner planet $i$ can be computed by swapping the subscripts in Equation \ref{eq:pot_pp}, in which case $n_{ji}$ becomes negative. Note that this expression contains both the static and time-dependent parts of the planet-forced tidal potential.

The potential arising from body $i$'s tidal deformation, commonly referred to as the response potential, is $\delta \Phi_i = k_2 \Phi_i$, where $k_2$ is the degree-2 potential tidal Love number \citep{love1911some}. We can use this expression to get some sense of how planet-planet tidal deformation compares to that from eccentricity tides, as was done by \citet{wright2018}. To order of magnitude, using Equations \ref{eq:pot_tides_ecc} and \ref{eq:pot_pp} evaluated at $\theta = \pi/2$, $\phi=0$ and $M=0$, we find;

\begin{equation}\label{eq:dr_ratio}
\frac{\delta \Phi_i^p}{\delta \Phi_i^e} \sim  \left(\frac{m_j}{m_\star}\right) \left(\frac{a_i}{r_{ij}}\right)^3 \frac{1}{3e_i}
\end{equation}

\noindent which reaches a maximum at planet $i$'s and $j$'s conjunction, $\text{min}(r_{ij}) = |a_i - a_j|$. Note that in this expression (which includes the factor of $1/3$ missing from \citet{wright2018}), we have assumed that $k_2$ is identical for both planet- and eccentricity-forcing, which is not the case in general. Equation \ref{eq:dr_ratio} is greatest for TRAPPIST-1 g due to tides from planet f, where the maximum tidal distortion may approach $\sim \SI{90}{\percent}$ of that from eccentricity-forcing (Table \ref{tb:params2}). This is illustrated in Figure \ref{fig:forcing_g}, which shows the magnitude of each forcing potential on TRAPPIST-1 g due to f as a function of time. For the other planets in the system except TRAPPIST-1 b, $\delta \Phi_i^p/\delta \Phi_i^e > 0.01$, and is sometimes greater than $0.1$, suggesting that in terms of deformation, planet-planet tides may be an important process in the system. This conclusion does not extend to tidal heating, though, where we also have to consider how the tidal forcing and planetary response changes with time.

In Figure \ref{fig:forcing_g} we see that planet-planet and eccentricity tides are temporally very different. The eccentricity-forcing on TRAPPIST-1 g operates at a single (orbital) frequency, while in contrast, tides due to planet f are a complex waveform composed of many frequencies. In order to evaluate tidal heating due to planet-forcing, we must decompose the forcing potential into each frequency component because a planet's tidal response is inherently frequency-dependent. This is the task of the next section.

\subsubsection{General form of the tidal potential}

The planet-planet tidal potential in Equation \ref{eq:pot_pp} can be expanded and written as an infinite sum of spherical harmonics; 

\begin{equation}\label{eq:pot_general}
\Phi^p(\theta, \phi, t) = \sum_{l=0}^{\infty} \sum_{m=-l}^{l} \Phi^p_{lm}(t) Y_{lm}(\theta, \phi).
\end{equation}

\noindent where $\Phi^p_{lm}$ are the degree-$l$ and order-$m$ cosine ($m\geq0$) and sine ($m<0$) spherical harmonic expansion coefficients of the potential $\Phi^p$, and the unnoramalised spherical harmonics are;
\begin{equation}
Y_{lm}(\theta, \phi) = 
\begin{cases}
P_{lm}(\theta)\cos(m\phi),& \text{if $m\geq 0$}\\
P_{l|m|}(\theta)\sin(|m|\phi),  &\text{if $m<0$}
\end{cases}
\end{equation}

\noindent where $P_{lm}(\theta)$ is the unnormalised associated Legendre function \citep{dahlen1998theoretical}. At each degree and order, $\Phi^p_{lm}(t)$ can be further expressed as an infinite sum over frequency, $q$;

\begin{equation}\label{eq:pot_pp_fourier}
\Phi^p_{lm}(t) = \frac{1}{2}a_{lm0} + \sum_{q=1}^{\infty} \left[ a_{lmq}\cos(qn_{ij}t) + b_{lmq} \sin(qn_{ij}t)\right]
\end{equation} 

\noindent where  $a_{lmq}$ and $b_{lmq}$ are the Fourier expansion coefficients of $\Phi_{lm}^p$ for each frequency $q$, as defined in Appendix \ref{ax:fourier}.

For each planet in the TRAPPIST-1 system, we calculate $a_{lmq}$ and $b_{lmq}$ by first evaluating the planet-planet tidal potential (Eq. \ref{eq:pot_pp}) over the planet's forcing period and surface. At each point in time, we then find $\Phi_{20}$, $\Phi_{22}$ and $\Phi_{2(-2)}$, giving us a time-series of the spherical harmonic expansion coefficients of the forcing potential (Appendix \ref{ax:sph}). Finally, a Fourier transform is applied to that time series to give us the coefficients of the Fourier series. Of these, the only nonzero Fourier series coefficients are $a_{20q}$, $a_{22q}$, and $b_{2(-2)q}$. We provide these coefficients for each planet in the TRAPPIST-1 system as supplementary material to this manuscript.

\begin{figure}[t]
	\centering
	\includegraphics[width=0.5\linewidth]{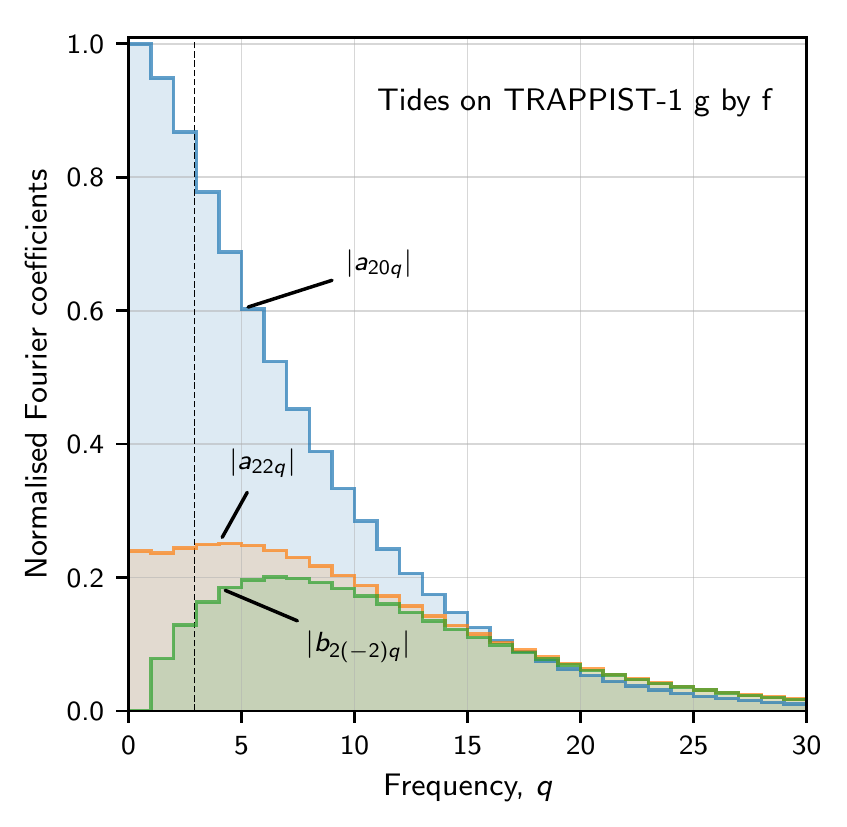}
	\caption{\label{fig:freq_spec} Normalised frequency spectrum of the three nonzero spherical harmonic coefficients of the planet tidal forcing potential for TRAPPIST-1 g due to f. The black dashed line shows the orbital frequency of TRAPPIST-1 g. Note that the equivalent frequency spectrum for the other TRAPPIST-1 planets are similar, although not identical, to that shown. The normalisation here is relative to the largest Fourier coefficient, $a_{20q}$.}
\end{figure}

As an example, Figure \ref{fig:freq_spec} shows the normalised frequency spectrum of these coefficients for the tidal potential on TRAPPIST-1 g due to planet f. What is particularly striking is that neither $a_{22q}$ or $b_{2(-2)q}$ peak at the conjunction frequency, $q=1$, but rather at higher frequencies. Additionally, both $a_{20q}$ and $a_{22q}$ have constant ($q=0$) components. Aside from the constant component, $a_{20q}$ is the only coefficient that peaks at the forcing frequency. Importantly, though, Figure \ref{fig:freq_spec} illustrates that planet-planet tides are composed of many frequencies, unlike the low-order eccentricity forcing, and much of the planet-planet forcing operates at frequencies higher than the orbital frequency. Consequently, when investigating the impact of planet-planet tides in the TRAPPIST-1 system, it is essential that we adopt a frequency-dependent approach.  

\subsubsection{The planetary response to tides}

As a planet deforms in response to tides, an additional gravitational potential arises as mass is redistributed around the body \citep{love1911some}. The potential arising from this deformation is commonly referred to as the response potential, $\delta\Phi$, and the magnitude and phase lag between this and the forcing potential controls the amount of tidal heating. The greater the planetary response, the larger $\delta\Phi$ becomes. For planet-planet tides, the spherical harmonic expansion coefficients of the response potential are;

\begin{equation}
\delta\Phi^p_{lm} = \Re\left\{\sum_{q=0}^{\infty} k_{l}(qn_{ij}) \left(a_{lmq} - i b_{lmq} \right) 
	\left[ \cos (qn_{ij}t) + i \sin(qn_{ij}t) \right] \right\}
\end{equation}

\noindent where the non-subscript $i$ is the imaginary number and $k_l(qn_{ij})$ is the complex degree-$l$ potential Love number \citep{love1911some} evaluated at the frequency $q n_{ij}$. The real part of the Love number $\Re(k_{l})$ represents the magnitude of the response potential, while the imaginary part $\Im(k_{l})$ corresponds to the phase lag between the forcing and response potentials. Explicitly writing $k_{l}$ in terms of its real and imaginary components, the response potential can be rewritten as;

\begin{equation}\label{eq:pot_resp}
\delta\Phi^p_{lm} = \sum_{q=0}^{\infty} \Re(k_{l}(qn_{ij})) \Bigl[ a_{lmq}\cos(qn_{ij}t) + b_{lmq}\sin(qn_{ij}t)\Bigr] + \Im(k_{l}(qn_{ij})) \Bigl[ b_{lmq}\cos(qn_{ij}t) - a_{lmq}\sin(qn_{ij}t)\Bigr] 
\end{equation}

In the following section we use this response potential and the forcing in Equation \ref{eq:pot_pp} to derive the tidal heating rate due to planet-planet tides.

\subsection{Tidal heating}\label{sec:heating}

Tidal heating in a solid body is generated via friction as the body responds through deformation to the external forcing potential. Unless the body is perfectly elastic, the response time is nonzero and lags by some amount behind the forcing potential. It is this lag that results in heating via friction at the microphysical scale. As previously mentioned, the lag is often characterized by the imaginary component of the tidal Love number, $k_l$, which depends on the forcing frequency for any anelastic material. In the case of planet-planet tides there are also multiple frequencies in the forcing itself (Figure \ref{fig:freq_spec}), and the tidal response will be different for each of those frequencies. A frequency-dependent approach is then essential when evaluating planet-planet tidal heating.

The time-averaged rate of energy dissipation over the forcing period $T_{ij} = 2\pi/n_{ij}$ is given in \citet[][Eq. 18]{zschau1978tidal};
\begin{equation}
\dot{E}^p_{ij} = -\frac{R_i n_{ij}}{8\pi^2 G} \sum_{l,m} (2l + 1) \int_{\Omega}\int_{0}^{T_{ij}} \delta\Phi^p_{lm} \left(\partial_t \Phi^p_{lm}\right) dt d\Omega,
\end{equation}

\noindent where $\partial_t$ represents a time-derivative, $d\Omega = \sin \theta d\theta d\phi$ is the solid angle, and we have collapsed the summation notation for convenience. 

If we substitute Equations \ref{eq:pot_pp_fourier} and \ref{eq:pot_resp} into the above and make use of the orthogonality conditions for the spherical harmonics $Y_{lmp}$ (Appendix \ref{ax:sph}), the time-averaged dissipation rate becomes;

\begin{equation}\label{eq:heating_pp}
\dot{E}^p_{ij} = \frac{R_i n_{ij}}{2 G} \sum_{l,m} \sum_{q=1}^{\infty} \frac{(l+|m|)!}{(l-|m|)!}\frac{1}{(2 - \delta_{m0})} q \Im(k_{l}(qn_{ij})) (a^2_{lmq} + b^2_{lmq})
\end{equation}

\noindent where $\delta_{lm}$ is the Kronecker delta function. We assume these bodies are incompressible, so tidal heating is unaffected by the static forcing and response potential, $q=0$,  and is consequently neglected in the frequency summation. For eccentricity-forcing the equivalent expression is \citep{segatz1988tidal};

\begin{equation}\label{eq:heating_ecc}
\dot{E}^e_i = -\frac{21}{2}\Im(k_{2}(n_i)) \frac{(n_i R_i)^2}{G}e_i^2
\end{equation}

\noindent which is limited to degree-2 and $k_2$ is evaluated at the orbital frequency, $n_i$. The Fourier series coefficients in Eq. \ref{eq:heating_pp}, $a_{lmq}$ and $b_{lmq}$, are computed from the planet-planet tidal potential (Eq. \ref{eq:pot_pp}) as described in the previous section. The only unknowns in this problem are then the frequency-dependent Love numbers for eccentricity- and planet-forcing, which we describe in the next section.

\section{Interior Structures and Rheology}\label{sec:rheology}

A planet's internal structure and rheology dictate the response of that planet to tides. There are sufficient uncertainties on the masses and radii of the TRAPPIST-1 planets that there is no great advantage in assuming complex interior structures. We instead choose to model these planets as homogeneous bodies with Maxwell rheologies in order to capture the planet's frequency-dependent response to tides. 

\begin{table}
	\centering
	\begin{tabular}{crrrrrrrr}
		\hline
		\multirow{ 2}{*}{Planet}&\multirow{ 2}{*}{$P$ [days]} &\multirow{ 2}{*}{$e$ [\num{e-2}]} &\multirow{ 2}{*}{$m$ [$M_\Earth$]} & \multirow{ 2}{*}{$R$ [$R_\Earth$]} & \multirow{ 2}{*}{$\dot{E}^p$ [\si{\watt}]} & \multirow{ 2}{*}{$\dot{E}^e$ [\si{\watt}]} & \multicolumn{2}{c}{$e_0$ [\num{e-2}]}\\
		 \cline{8-9}  &  & & &  &  &  & $\eta_{low}$ &  $\eta_{high}$ \\\hline\hline
		b & 1.519 & $0.622 \pm 0.304$ & $1.017 \substack{ +0.154 \\ -0.143}$ & $1.121 \substack{ +0.031 \\ -0.032}$ & \num{ 8.1e+07 } & \num{ 9.7e+11 } & 0.014 & 0.006 \\
		c & 2.435 & $0.654 \pm 0.188$ & $1.156 \substack{ +0.142 \\ -0.131}$ & $1.095 \substack{ +0.030 \\ -0.031}$ & \num{ 5.9e+07 } & \num{ 1.6e+11 } & 0.039 & 0.013 \\
		d & 4.072 & $0.837 \pm 0.093$ & $0.297 \substack{ +0.039 \\ -0.035}$ & $0.784 \substack{ +0.023 \\ -0.023}$ & \num{ 8.5e+05 } & \num{ 3.2e+09 } & 0.041 & 0.014 \\
		e & 6.135 & $0.510 \pm 0.058$ & $0.772 \substack{ +0.079 \\ -0.075}$ & $0.910 \substack{ +0.026 \\ -0.027}$ & \num{ 3.3e+05 } & \num{ 9.4e+08 } & 0.028 & 0.009 \\
		f & 9.261 & $1.007 \pm 0.068$ & $0.934 \substack{ +0.080 \\ -0.078}$ & $1.046 \substack{ +0.029 \\ -0.030}$ & \num{ 8.9e+05 } & \num{ 1.4e+09 } & 0.069 & 0.025 \\
		g & 12.426 & $0.208 \pm 0.058$ & $1.148 \substack{ +0.098 \\ -0.095}$ & $1.148 \substack{ +0.032 \\ -0.033}$ & \num{ 7.1e+05 } & \num{ 2.9e+07 } & 0.096 & 0.032 \\
		h & 18.871 & $0.567 \pm 0.121$ & $0.331 \substack{ +0.056 \\ -0.049}$ & $0.773 \substack{ +0.026 \\ -0.027}$ & \num{ 4.1e+03 } & \num{ 3.6e+06 } & 0.059 & 0.019 \\
		\hline
	\end{tabular}
	\caption{Relevant geophysical parameters, orbital parameters, are tidal heating results for the TRAPPIST-1 planets. Eccentricity, $e$, semimajor axis, $a$, and the masses, $m$, are from \citet{grimm2018nature}, while the orbital periods, $P$ and planetary radii, $R$, are from \citet{delrez2018early}. Uncertainties are shown to $1\sigma$, and we include uncertainties for the most unconstrained parameters. We also use a stellar mass of $0.089\pm0.007 M_\Sun$ in this work \citep{delrez2018early}. Tidal heating from planet-planet tides, $\dot{E}^p$ (Eq. \ref{eq:heating_pp}), and eccentricity tides, $\dot{E}^e$ (Eq. \ref{eq:heating_ecc}), are shown for a nominal viscosity of \SI{e21}{\pascal\second}. The last two columns are estimates of how small the orbital eccentricity must be if tidal heating due to planet-forcing is to become comparable to eccentricity-forced heating, $e_0$, for both a low ($\eta_{low} < \SI{e14}{\pascal\second}$) and high viscosity ($\eta_{high} > \SI{e16}{\pascal\second}$) Maxwell body (Figs. \ref{fig:trappist_heating} and \ref{fig:heating_ecc}).\label{tb:params2}}
\end{table}

The two properties that control the behaviour of a Maxwell material are its viscosity, $\eta$, and rigidity, $\mu$. The Maxwell time, $\tau_M = \eta/\mu$, is a fundamental property of such a material, and gives the transition timescale from elastic to viscous deformation. If a Maxwell material is forced on a timescale less than $\tau_M$, the material response is elastic. If the opposite is true, then the response is viscous. For average silicate planetary properties of $\mu=\SI{50}{\giga\pascal}$ and $\eta=\SI{e21}{\pascal\second}$ \citep{henning2009tidally}, $\tau_M >$ 600 years. The forcing period from either eccentricity or planet-planet tides in the TRAPPIST-1 system is much shorter than this Maxwell time (Table \ref{tb:params2}), so we would expect the planet response to be elastic. We note, however, that the forcing period can approach the Maxwell time for low viscosities ($\eta \sim \SI{e14}{\pascal\second}$). We choose a Maxwell rheological model in this work because it is the most simple method to capture the frequency-dependent response of a material.

For a homogeneous Maxwell body, the imaginary part of the degree-2 potential Love number at frequency $qn_{ij}$ can be expressed analytically as \citep[e.g.,][]{henning2009tidally, renaud2018increased},

\begin{equation}
\Im(k_{2}(qn_{ij})) = -k_f 
\left[\frac{q\bar{\mu}n_{ij}\tau_M}{1 + \left(q(1+\bar{\mu})n_{ij}\tau_M\right)^2} \right],\label{eq:k2_imag}
\end{equation}

\noindent where $\bar{\mu} = (19/2)\mu/\rho g R$ is the effective rigidity of the homogeneous body with surface gravity $g$, and $k_f = 3/2$ is the fluid degree-2 Love number \citep{love1911some}. This expression is sensitive to frequency, and this is illustrated in Appendix \ref{ax:love}. Using Equation \ref{eq:k2_imag}, we calculate $\Im(k_{2})$ up to frequency $q=100$ in Eq. \ref{eq:heating_pp} to evaluate tidal heating due to planet-planet tides.

\section{Results and Discussion}\label{sec:results}

\begin{figure}[t]
	\centering
	\includegraphics[width=\linewidth]{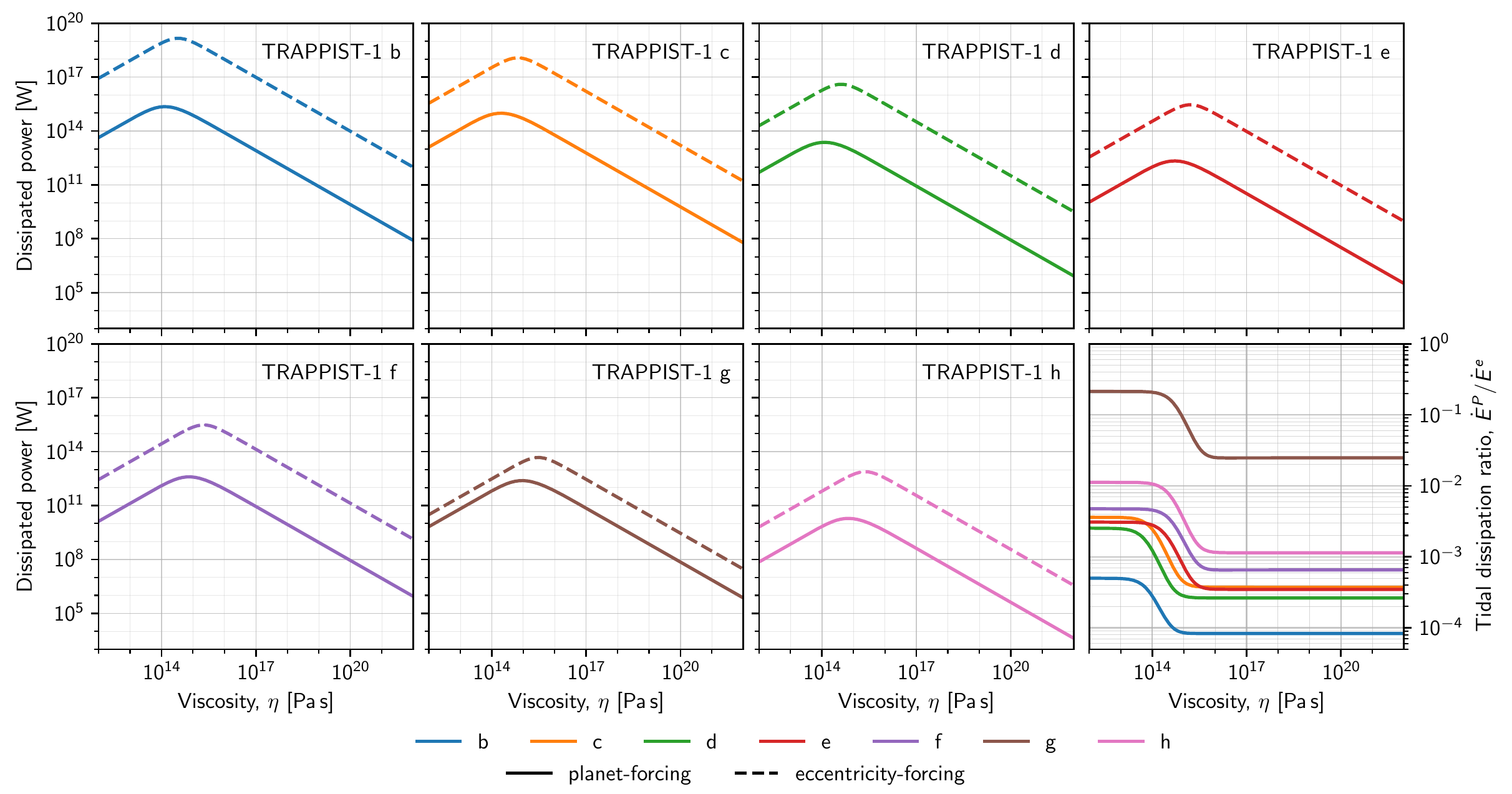}
	\caption{Tidally dissipated power for each TRAPPIST-1 planet as a function of viscosity, assuming $\mu=\SI{50}{\giga\pascal}$ and a homogeneous Maxwell rheology. Dashed lines show dissipated power from our model using eccentricity forcing (Eq. \ref{eq:heating_ecc}), while the solid lines are calculated using planet-forcing (Eq. \ref{eq:heating_pp}) from all the planets. The bottom right panel shows the ratio of planet- to eccentricity-forced tidal heating for each planet.}\label{fig:trappist_heating}
\end{figure}

The primary aim of this manuscript is to identify how significant planet-forced tidal heating can be relative to eccentricity forcing. The most significant unknown in this problem is the bulk viscosity of the material, $\eta$, which we vary over several orders of magnitude. Figure \ref{fig:trappist_heating} shows the tidal dissipated power for each TRAPPIST-1 planet as a function of their bulk viscosities. The last panel shows the ratio of dissipated power from planet-planet tides to eccentricity tides for each planet.

Most notably, Figure \ref{fig:trappist_heating} shows that tidal heating from planet forcing is always less than eccentricity forcing over the explored viscosity parameter space. This fact does not change over a range of reasonable rigidities, or with different rheologies (Appendix \ref{ax:andrade}). We can then conclude, using the eccentricities in Table \ref{tb:params2} from \citet{grimm2018nature}, that planet-planet tidal heating is likely secondary to stellar eccentricity tides. TRAPPIST-1 g is the planet that appears to experience the most significant amount of planet-planet tidal heating, relative to eccentricity forcing.

Interestingly, there are two main solutions to the tidal heating ratio,  one in a high viscosity regime ($\eta \gtrsim $ \SI{e16}{\pascal\second}) and the other in a much lower viscosity regime ($\eta \lesssim \SI{e14}{\pascal\second}$), as highlighted in the bottom right panel of Figure \ref{fig:trappist_heating}. For all planets, transition from the high to low viscosity regime increases the ratio of planet to eccentricity tidal heating by around one order of magnitude. For TRAPPIST-1 g, this corresponds to planet-planet tides accounting for as low as \SI{2}{\percent} and up to \SI{20}{\percent} of the total amount of tidal heating from the high to low viscosity regime, respectively. TRAPPIST-1 h has the second most significant amount of planet-planet tidal heating, which accounts for up to \SI{1}{\percent} of the total amount of tidal heating, but only for the low viscosity regime. Planet-planet tides are negligible for all other planets, where they account for $<\SI{1}{\percent}$ of the total amount of tidal heating.

Based on our comparison of the maximum tidal deformation in Section \ref{sec:theory} (Eq. \ref{eq:dr_ratio}), it is surprising that tidal heating from planet-planet tides is in general so much less than that from stellar eccentricity tides. This is especially true on TRAPPIST-1 g, where the maximum tidal deformation from planet f is nearly equal to that from eccentricity tides. The reason for this difference is because the planet responds far more elastically to planet-forcing than it does to eccentricity-forcing. Much of the power in the frequency spectrum for planet-forcing is at frequencies higher than the orbital frequency (Fig. \ref{fig:freq_spec}). At these high frequencies, the imaginary part of $k_{2}$ becomes very small, meaning that the planet responds more elastically to the forcing (Appendix \ref{ax:love}). The more elastic the response is, the less internal friction there is and consequently tidal dissipation drops.    


The peaks in tidal heating shown in Figure \ref{fig:trappist_heating} all occur because, for those viscosities, the Maxwell time becomes comparable to the forcing period. Eccentricity and planet-planet tides have different forcing periods, meaning they each have different viscosities where the Maxwell time approaches the period of forcing. This is why for each planet, the maximum tidal heating occurs at different viscosities for eccentricity and planet-forcing. As a consequence, there is a transition in the tidal heating ratio as the viscosity changes. As discussed above, a planet responds more elastically to planet-planet tides than it does to eccentricity tides in general, which is why the peak in tidal heating occurs at smaller viscosities for planet-forcing than it does for eccentricity-forcing. We observe a similar tidal heating ratio transition for other rheologies (Appendix \ref{ax:andrade}).

\subsection{Importance of orbital eccentricity}

\begin{figure}[t]
	\centering
	\includegraphics[width=0.8\linewidth]{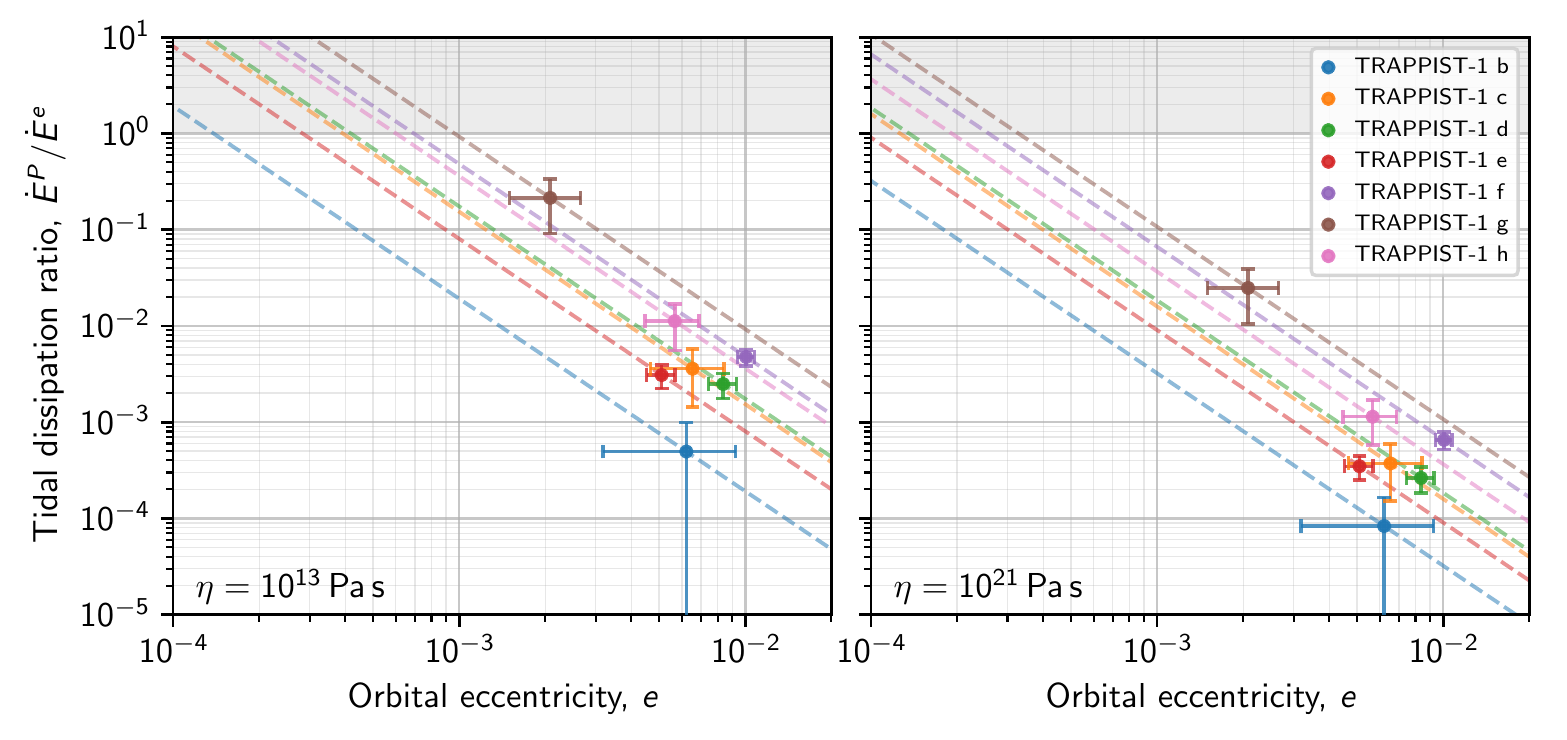}
	\caption{Ratio of tidal dissipation from planet- to eccentricity-forcing for each TRAPPIST-1 planet as a function of orbital eccentricity, assuming $\mu=\SI{50}{\giga\pascal}$. Dashed lines show how this ratio varies with changing eccentricity. The left and right panels are for a low (\SI{e13}{\pascal\second}) and high (\SI{e21}{\pascal\second}) viscosity scenario, respectively. Error bars represent 1-sigma uncertainties in only the eccentricity heating, based on the uncertainties in planetary mass, radius, and eccentricity (Table \ref{tb:params2}). The grey region represents where planet-forced tidal heating is comparable to or greater than that due to stellar eccentricity tides. Eccentricities are from \citet{grimm2018nature}. The eccentricities that correspond to a tidal heating ratio of unity, $e_0$, are given in Table \ref{tb:params2}.}\label{fig:heating_ecc}
\end{figure}

The orbital eccentricities of the TRAPPIST-1 planets are a fundamental property in this problem (Eq. \ref{eq:heating_ecc}). While the most current eccentricities are reasonably well constrained \citep{grimm2018nature}, we caution that the eccentricity and mass of a planet can be correlated \citep[e.g.,][]{lithwick2012extracting}. Additionally, the eccentricities of the planets are likely to change significantly over short timescales \citep[e.g.,][]{luger2017seven} due to both tidal damping and perturbations from the other planets. Due to this, we also explore how changes in eccentricity affect the planet to eccentricity-forced tidal heating ratio, shown in Figure \ref{fig:heating_ecc}. The dashed lines in Figure \ref{fig:heating_ecc} represent possible solutions for the planet to eccentricity-forced tidal heating ratio. We explore a low (left) and high (right) viscosity solution. 

None of the TRAPPIST-1 planets lie inside the gray region, which is where planet-planet heating is comparable to or greater than eccentricity heating. For a low viscosity of $\eta=\SI{e13}{\pascal\second}$, TRAPPIST-1 g need only halve its eccentricity to create equal parts planet and eccentricity forced tidal heating. The eccentricities of the other planets must be reduced from the \citet{grimm2018nature} values (Table \ref{tb:params2}) by an order of magnitude or more for this to happen. Planet-planet heating is negligible relative to eccentricity heating for a viscosity of $\eta = \SI{e21}{\pascal\second}$. We conclude that for the majority of the TRAPPIST-1 planets, their eccentricities need to be \numrange{1}{2} orders of magnitude lower than those predicted in \citet{grimm2018nature} for planet-planet tidal heating to become at least as significant as eccentricity-driven stellar tidal heating. The exact eccentricities that correspond to a tidal heating ratio of unity, $e_0$, are given in Table \ref{tb:params2}. For most planets this corresponds to $e_0 = $ \numrange{e-3}{e-4}. TRAPPIST-1 g is the only planet where planet-forced tidal heating accounts for $>\SI{1}{\percent}$ of the total amount of tidal heating throughout the explored viscosity parameter space. Up to $\SI{20}{\percent}$ of all tidal heating on TRAPPIST-1 g can come from planet-forced tides in the low viscosity regime (assuming $e=\num{0.0051}$ \citep{grimm2018nature}). 

\subsection{Orbital evolution}

Tidal heating due to orbital eccentricity reduces the eccentricity of a planet as well as slightly shrinking its semimajor axis \citep{murray1999solar}. Planet-planet tides, which exist even in circular orbits, can only shrink the semimajor axis of the tidally heated body. Depending on the relative speed of inward migration between two planets tidally heating each other, an unstable ``pile-up'' may occur.  While possible, we find that the characteristic semimajor axis decay timescale for all TRAPPIST-1 planets, $\tau_i = a_i/\dot{a}_i = Gm_\star m_i/ 2 a_i \dot{E}_{ij}^P$ \citep{murray1999solar}, is $>\SI{100}{\giga yr}$ because these planets are so close to their host star. We therefore conclude that planet-planet tides likely have a negligible effect on the orbital evolution of the TRAPPIST-1 planets. We also note that torques from neighboring planets may impact the rotation rate of the tidally distorted planet, but that is beyond the scope of this work.

\subsection{Caveats}

The absolute amount of tidal dissipation due to planet-forcing is very much of interest, but we caution that this strongly depends on the assumed rheological model and interior structure of the body. Maxwellian rheologies can greatly underestimate the amount of possible tidal dissipation, as has been demonstrated for the Jovian satellite Io \citep{bierson2016test,renaud2018increased}. For that reason we do not focus on the absolute amount of tidal heating, but rather the heating ratio between these two modes of tidal forcing. In Appendix \ref{ax:andrade} we present an equivalent version of Figure \ref{fig:trappist_heating} that uses an Andrade rheology \citep{andrade1910viscous, jackson2010grainsize}, which produces much greater amounts of tidal heating from both eccentricity- and planet-forcing in the high viscosity regime. The ratio of planet to eccentricity forced tidal heating in the high viscosity regime also increases slightly for all planets, but this is rather parameter dependent. Furthermore, most rheological models and their associated parameters are derived from laboratory experiments under constant or simple periodic forcing conditions \citep[e.g.,][]{jackson2004shear, jackson2010grainsize}. While certainly periodic, planet-forcing contains multiple frequency components (Figure \ref{fig:freq_spec}). Ideally, a rheological model applied to planet-planet tides should therefore be based on laboratory work under the same forcing conditions, which to our knowledge has not yet been performed.

An implicit assumption in our derived tidal heating rate (Eq. \ref{eq:heating_pp}) is that the system responds linearly to the forcing. In other words, the total response of a planet to tides can be given by the summation of the response at each individual frequency. For the inner two TRAPPIST-1 planets, where planet-planet tides are the most extreme, this assumption should be taken with caution.

Finally, we ignore fluid tides in this work for simplicity. It has been suggested that some of the TRAPPIST-1 planets may have low density envelopes comprising up to \SI{5}{\percent} their mass \citep{grimm2018nature}. If these envelopes are primarily liquid, they may respond more strongly to planet-forced tides than the solid body because dynamical ocean tides have higher natural frequencies in their response \citep[e.g.,][]{kamata2015tidal, hay2019nonlinear}.


\section{Conclusions}

The importance of tidal heating due to tides raised by one planet on another in the TRAPPIST-1 system is investigated in this manuscript. To do this, we derive the tidal potential on a planet due to the gravitational attraction of its neighbours. The potential, which contains may high frequency components, is decomposed first into spherical harmonics and then into Fourier coefficients. We calculate tidal heating on each planet due to tides from every other planet in the TRAPPIST-1 system using these Fourier coefficients and assuming homogeneous Maxwell material interiors. The amount of tidal heating due to neighboring planets is compared to eccentricity-forced tidal heating from the star.

Planet-planet tidal heating is found to always be less than that due to eccentricity tides (assuming orbital eccentricities from \citet{grimm2018nature}), because planets respond more elastically to planet-forcing than they do to eccentricity-forcing due to high frequency components in the planet-planet tidal potential. For low viscosities ($<\SI{e14}{\pascal\second}$), planet-planet tidal heating is around an order of magnitude more significant than if the viscosity were high ($>\SI{e16}{\pascal\second}$), although eccentricity-forced tides still dominate. This transition in tidal heating ratio from a high to low viscosity is because planet-forced tides operate at the planetary conjunction frequency (and higher), while eccentricity-forcing only occurs at the orbital frequency.

Viscosities lower than \SI{e14}{\pascal\second} seem unlikely unless the body is largely icy or has an extremely high degree of partial melt \citep[e.g.,][]{bierson2016test,barr2018interior,renaud2018increased}. We therefore prefer a high viscosity scenario for these planets, which lowers the effectiveness of planet-forced tidal heating relative to eccentricity-forcing. Adopting different rheological models, such as the Andrade rheology, may improve this outlook (Appendix \ref{ax:andrade}). 

Changes in orbital eccentricity also have a strong effect on the relative importance of planet-forced tidal heating. For TRAPPIST-1 g, its eccentricity from \citet{grimm2018nature} must halve for tidal heating from eccentricity- and planet-forcing to become equal. For the other planets, an eccentricity drop of between \numrange{1}{2} orders of magnitude is needed for planet-forced heating to become equal to eccentricity-forced heating. 

Overall, planet-planet tidal heating is not found to be significant in the TRAPPIST-1 system when compared to tidal heating due to orbital eccentricity, except for TRAPPIST-1 g. This conclusion is based on our assumption of a homogeneous solid Maxwell body. We neglect any dynamical fluid tides in our model, which may be a source of additional tidal heating and an avenue for future research.

\section{Acknowledgments}

We thank Renu Malhotra for insightful and encouraging discussions, and Jessie Brown for detailed comments on the manuscript. Some of the ideas presented in this manuscript were formulated through discussion with the participants of the 2018 Keck Institute for Space Studies workshop on tidal heating. This work was supported by the National Aeronautical and Space Agency (NASA) through the Habitable Worlds program (NNH15ZDA001N-HW).

\appendix

\section{Fourier Expansion}\label{ax:fourier}

We decompose each spherical harmonic coefficient of the tidal potential, $\Phi^p_{lm}$ (Eq. \ref{eq:pot_general}), into Fourier series coefficients (Section \ref{sec:theory}). We define the degree-$l$ and order-$m$ Fourier series coefficients at frequency $q$ as;

\begin{align}
a_{lmq} &= \frac{2}{T_{ij}} \int_{0}^{T_{ij}} \Phi^p_{lm}(t) \cos(qn_{ij}t) dt \\
b_{lmq} &= \frac{2}{T_{ij}} \int_{0}^{T_{ij}} \Phi^p_{lm}(t) \sin(qn_{ij}t) dt
\end{align}

\noindent where $T_{ij} = 2\pi /n_{ij}$ is the forcing/conjunction period. These integrals are evaluated using a Discrete Fast Fourier Transform from the Python library numpy.

\section{Spherical Harmonics}\label{ax:sph}

\subsection{Spherical harmonic coefficients}

The tidal potential due to planet forcing (Eq. \ref{eq:pot_pp}) is evaluated over the reference radius of the deformed body and is then decomposed into spherical harmonics at each point in time. To find the spherical harmonic expansion coefficients at each degree-$l$ and order-$m$, we use the FORTRAN-95 library shtools \citep{wieczorek2018shtools}.

\subsection{Orthogonality}

Spherical harmonics are orthogonal over all degrees $l$ and orders $m$ with the normalisation \citep[e.g.,][]{wieczorek2015gravity};

\begin{equation}
	\int_{\Omega} Y_{lm} Y_{l'm'}d\Omega = 
	\frac{4\pi \delta_{ll'}\delta_{mm'} }{2l + 1}\frac{(l + |m|)!}{(l - |m|)!} \frac{1}{(2 - \delta_{m0})}
\end{equation}

\noindent where $d\Omega=\sin \theta d \theta d\phi$ is the solid angle and $\delta_{lm}$ is the Kronecker delta function. This condition is used to derive the energy dissipation expression in Eq. \ref{eq:heating_pp}.

\section{Love numbers of a Maxwell material}\label{ax:love}

\begin{figure}[t]
	\centering
	\includegraphics[width=0.5\linewidth]{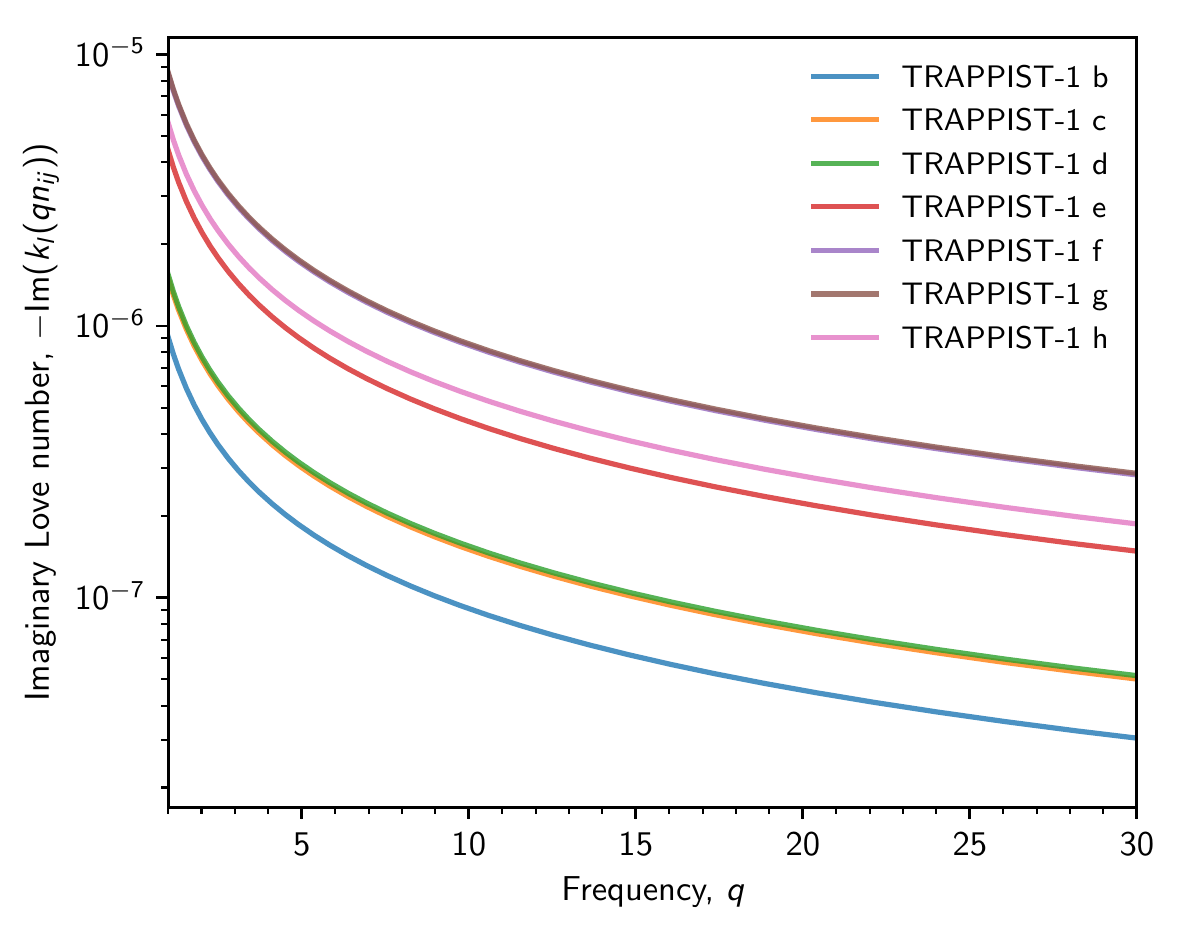}
	\caption{The imaginary component of the degree-2 tidal Love number as a function of frequency, $q$. The frequency shown is relative to the conjunction frequency with each planet's nearest neighbour, $qn_{ij}$.}\label{fig:k2}
\end{figure}

The imaginary component of the degree-2 tidal Love number, $\Im (k_{2})$, varies strongly with frequency. We show the $\Im (k_{2})$ of a Maxwell material (Eq. \ref{eq:k2_imag}) for each TRAPPIST-1 planet as a function of frequency in Figure \ref{fig:k2}, where the frequency shown is relative to the conjunction frequency with each planet's nearest neighbour. 

The imaginary part of $k_{2}$ decreases rapidly with increasing frequency. For planet-planet tides, which contain high frequency components, this means that the $\Im(k_{2})$ that corresponds to the average frequency in the forcing is much smaller than it is for the conjunction frequency. For this reason, planet-planet tidal heating suffers because the dominant frequency in the forcing corresponds to a smaller $\Im(k_{2})$.

\section{Andrade Rheology}\label{ax:andrade}

\begin{figure}[t]
	\centering
	\includegraphics[width=\linewidth]{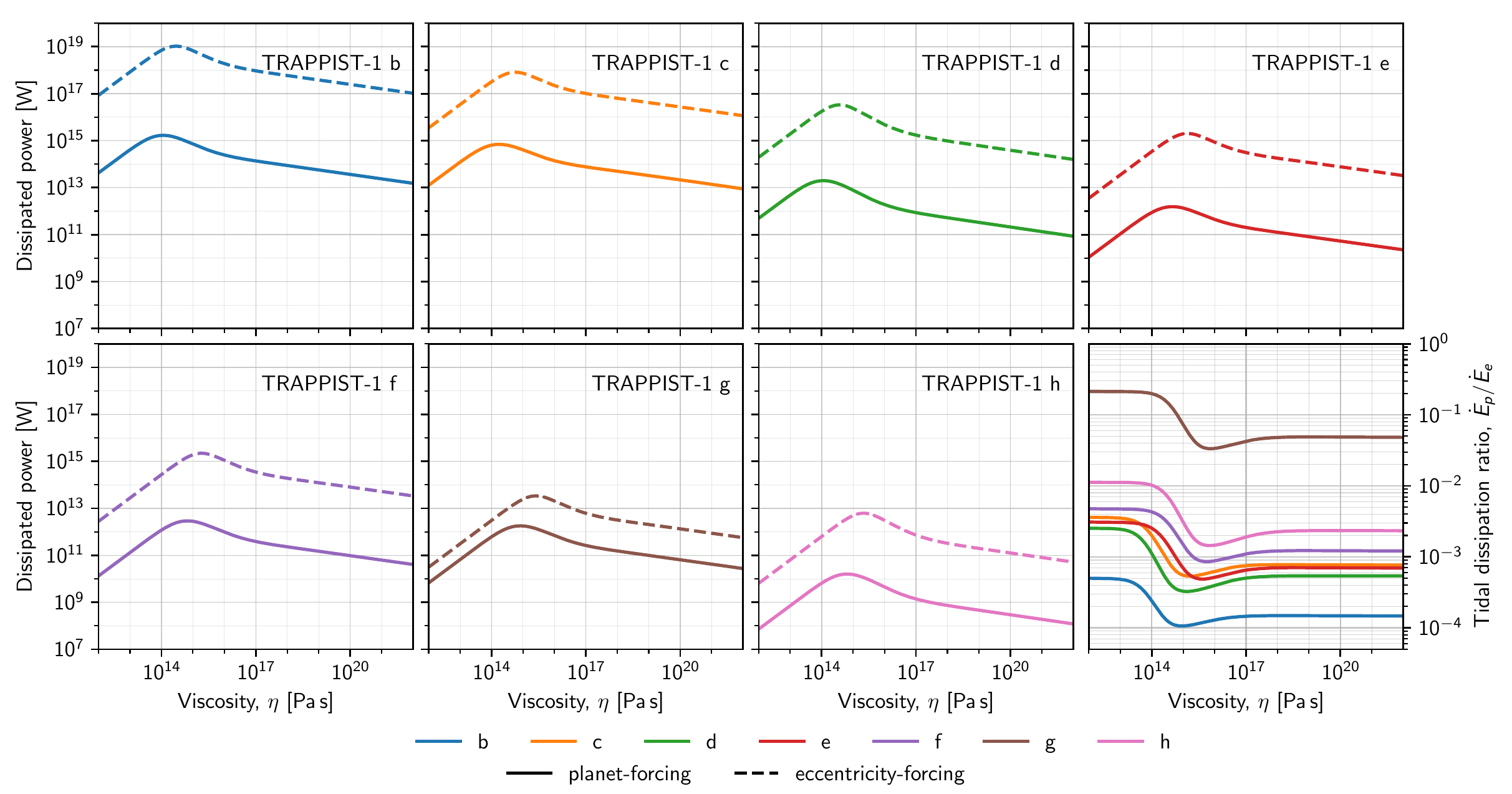}
	\caption{Tidally dissipated power for each TRAPPIST-1 planet as a function of viscosity, assuming $\mu=\SI{50}{\giga\pascal}$ and using an Andrade rheology \citep{renaud2018increased}. The solid lines are calculated using planet-forcing (Eq. \ref{eq:heating_pp}) and the dashed lines are for eccentricity-forcing (Eq. \ref{eq:heating_ecc}). The bottom right panel shows the ratio of planet- to eccentricity-forced tidal heating for each planet.}\label{fig:trappist_heating_andrade}
\end{figure}

Most anelastic materials do not behave in a Maxwellian fashion. A more advanced rheological model that is better suited for tides is the Andrade rheology \citep[e.g.,][]{andrade1910viscous, jackson2010grainsize}. Recently, \citet{renaud2018increased} showed that tidal dissipation could be much stronger in high viscosity materials when using an Andrade rheology instead of a Maxwell approach. Unfortunately, modeling Andrade materials uses several more free parameters than a Maxwell material does, and some of these are difficult properties to measure in laboratory experiments. The imaginary part of the degree-2 Love number for a homogeneous Andrade body is \citep[e.g.,][Table 3]{renaud2018increased};

\begin{equation}
\Im(k_{2}(qn_{ij})) = -k_f 
\left[\frac{\tau_M q n_{ij} \bar{\mu} [1 + (\tau_M q n_{ij})^{1-\alpha} \zeta^{-\alpha}S]}
{1 +
 (\tau_M q n_{ij})^2(\bar{\mu} + 1)^2 +
 (\tau_M q n_{ij})^{2(1-\alpha)} \zeta^{-2\alpha} (\alpha !)^2 +
 2(\tau_M q n_{ij})^{2-\alpha} \zeta^{-\alpha}
 [S/(\tau_M q n_{ij}) + (\bar{\mu} + 1)C]
} \right],\label{eq:k2_imag_andrade}
\end{equation} 

\noindent where $\alpha$ and $\zeta$ are the Andrade empirical exponent and timescale, respectively, and the Andrade constants are;

\begin{align}
S &= \alpha! \sin (\alpha \pi / 2), \\
C &= \alpha! \cos (\alpha \pi / 2).
\end{align} 

Using nominal values for the empirical constants from \citet[Table 1]{renaud2018increased}, $\alpha = 0.8$ and $\zeta = 1$, we calculate $\Im(k_{2})$ using \ref{eq:k2_imag_andrade} for planet-planet and eccentricity tides to estimate the resulting tidal dissipation (Eqs. \ref{eq:heating_pp}, \ref{eq:heating_ecc}), shown in Figure \ref{fig:trappist_heating_andrade}. 

We note two significant differences between the Maxwell (Fig. \ref{fig:trappist_heating}) and Andrade models. Most significantly, the amount of tidal dissipation has increased by several orders of magnitude for both planet- and eccentricity-forcing. Secondly, the different shape of the dissipation curves results in slightly higher ratios between planet-planet and eccentricity tidal heating when the viscosity is $> \SI{e17}{\pascal\second}$. These results are sensitive to both $\alpha$ and $\zeta$, and this could be explored in future work.

\bibliography{mybibfile}



\end{document}